\documentstyle[preprint,revtex]{aps}
\begin{document}
\preprint{WIS--93/96/Sept.--PH}
\draft
\begin{title}
Single Level Current and Curvature Distributions \\
in Mesoscopic Systems.
\end{title}
\author{Alex Kamenev$^1$ and Daniel Braun$^2$}
\begin{instit}
$^1$Department of Nuclear Physics, The Weizmann Institute of Science,
Rehovot 76100, Israel.\\
$^2$Laboratoire de Physique des Solides, Associe an CNRS, Bat 510,
Universite  Paris--Sud, 91450, Orsay, France.
\end{instit}
\begin{abstract}
Exact analytic results for single level current and curvature
distribution functions  are derived in a framework of a
$2\times 2$ random matrix model. Current and curvature are
defined as the first and second derivatives of energy with respect to
a time--reversal symmetry breaking parameter (magnetic flux). The
applicability of the obtained distributions for the spectral statistic of
disordered
metals is discussed. The most surprising feature of our results is the
divergence of the
second and higher moments of the curvature at zero flux. It is shown that this
divergence  also appears in the general $N\times N$ random matrix model.
Furthermore, we find
an unusual logarithmic behavior of the two point current--current correlation
function at small flux.

\end{abstract}



\section{Introduction}
\label{s1}

The statistical properties of a single electron energy spectra in disordered
mesoscopic systems have  been the  subject of  intensive study during the
past decade. Apart from a fundamental interest on it's own, the spectral
statistic is closely related to such observable characteristics as
persistent currents, anomalous magnetization, and conductance fluctuations.
After
the seminal studies of Efetov \cite{Efetov83} and Altshuler and Shklovski\u{\i}
\cite{Altshuler86} it became clear that the spectral statistic in the diffusive
regime, where the mean free path $l$ of the electrons is much smaller than the
sample size $L$, may be
described by  random matrix theory (RMT).
The major feature of the spectra, known long ago from RMT
\cite{Forster64,Mehta91}, is  a level
repulsion. It means that the probability, $P(\epsilon)$, to find two
consecutive levels a distance $\epsilon$ apart tends to zero as $\epsilon$
decreases: $P(\epsilon)\stackrel{\epsilon\rightarrow 0}{\longrightarrow} 0$.
Despite  great progress in the theory,  closed analytical expressions for
the distribution function (DF), $P(\epsilon)$, are  not available (except
small $\epsilon$ behavior). It was demonstrated, however, in a vast amount
of numerical experiments \cite{Mehta91}, that a simple Wigner surmise,
obtained for a
$2\times 2$ random Hamiltonian, is an excellent approximation of the
$P(\epsilon)$ of  large
$N\times N$ matrices. Originally the Wigner DF was
obtained for three distinct symmetry classes of the Hamiltonian: Gaussian
orthogonal,
unitary and symplectic ensembles ((GOE),(GUE) and (GSE), respcectively)
\cite{Forster64}.
Subsequently the
crossover ensembles from one pure symmetry type to another were introduced and
investigated \cite{Pandey83}. In the present study we are especially
interested in a
crossover from GOE to GUE, which corresponds to a gradual breaking of
the time--reversal symmetry. To this end it was proposed \cite{Pandey83} to
study a random Hamiltonian of the following structure:
\begin{equation}
H^{(N)}(\alpha)=H^{(N)}_S+i\alpha H^{(N)}_A,
\hskip 2cm         0\leq \alpha \leq 1,
                                                            \label{ham}
\end{equation}
where $H^{(N)}_S$ and  $H^{(N)}_A$ are, respectively, symmetric and
antisymmetric
real random \mbox{$N\times N$} matrices. For $\alpha=0$ one has GOE,
for
$\alpha=1$ the  GUE--case. It was shown recently \cite{Montambaux91} that the
time--reversal symmetry breaking parameter $\alpha$ may be uniquely connected
to an Aharonov--Bohm flux, $\Phi\equiv \phi\frac{\Phi_0}{2\pi}$,
carried by a flux line, which penetrates  the system.
The latter, by gauge invariance, may be related to a change of boundary
conditions, imposed on wave functions in the angular direction (around the
flux line): $\Psi(2\pi)=\Psi(0) e^{i\phi}$ \cite{Bayers63}. According to Ref.\
\cite{Montambaux91} the relation between $\alpha$ and $\phi$ has the form
\begin{equation}
\alpha=\sqrt{\frac{\pi E_c}{N\Delta}}\phi,
                                                            \label{conj}
\end{equation}
where $\Delta$ is the mean level spacing  and $E_c$ is the Thouless
correlation energy of the system.

Let us now point out that the first and second derivatives of the
energy levels with respect to flux (hence with respect to $\alpha$) have
a clear physical interpretation. The first derivative
$\partial\epsilon_n/\partial\Phi$ is exactly  a single level current, carried
by each energy level  in the absence of time reversal symmetry
($\alpha>0$). These single
level currents manifest themselves, for example, in persistent currents through
a mesoscopic ring. As was conjectured by Thouless \cite{Thouless77} and
subsequently discussed by Akkermans and Montambaux \cite{Akkermans91},
a typical single level curvature (the second derivative with respect to the
flux)
may be considered as a measure of the correlation energy, hence it is directly
proportional to the dissipative conductance of the system. The  Thouless
conjecture is usually written as \cite{Montambaux91,Akkermans91}
\begin{equation}
E_c=\left[\langle\left(\left.\frac{\partial^2\epsilon_n}{\partial \phi^2}
\right|_{\phi=0}\right)^2\rangle\right]^{1/2},
                                                          \label{second}
\end{equation}
where the angular brackets $\langle\ldots\rangle$ denote an averaging over
a random ensemble. It  certainly makes sense to ask the question, what are
the distributions of the single level current and curvature, as a functions of
flux, in the framework of RMT. Today one knows only the lowest correlators of
these quantities like an average value,  variance \cite{Oppen91,Gefen91} or
two point pair
correlator (see e.g. Ref.\ \cite{Altshuler93}). This knowledge comes mainly
from  diagrammatic perturbation theory, which is usually not applicable
to small values of flux.

The aim of the present article is to derive  exact analytical results for the
single level current and curvature DF's, using RMT.
We shall also find all the moments of these quantities  and compare them
with available perturbative results. Finally we shall
calculate  the two point current--current correlation function for a small
flux.
As  was done in the case of the level spacing distribution,
we shall use a simple but exactly solvable $2\times 2$ random matrix model. We
prove, however, that some key features of our results are not restricted to
this toy model but apply to  $N\times N$ random matrices as well.

The most surprising features resulting  from the present study are the
following: Let us define a  single level curvature to be proportional to the
second flux derivative of energy
$$
c\sim\frac{1}{E_c}
\left|\frac{\partial^2\epsilon_n}{\partial \phi^2}\right|$$
(numerical factors will follow in the body of the article). Then the flux
dependent distribution function of the curvature has the form (for
$\phi\ll\phi_c\equiv\sqrt{\Delta/(2E_c)}\ll 1$)
\begin{equation}
P_{\phi}(c) \sim  \left\{ \begin{array}{ll}
{\displaystyle c^{-1/2} ;}
& c\ll 1,  \\
{\displaystyle  c^{-3} ;}
& {\displaystyle 1\ll c \ll \phi_c/\phi,} \\
{\displaystyle c \frac{\phi^4}{\phi_c^4}
\exp\left(-\frac{\phi^2}{\phi_c^2} c^2\right) ;  } \hskip .5cm
& {\displaystyle \phi_c/\phi \ll c }
\end{array} \right.
                                                          \label{new1}
\end{equation}
This distribution implies the following flux dependence of the
moments of the curvature
\begin{equation}
\langle \left|\frac{\partial^2\epsilon_n}{\partial \phi^2}\right|^m \rangle
\sim \left\{ \begin{array}{ll}
{\displaystyle E_c;}
& m=1, \\
{\displaystyle E_c^2 \ln(\phi_c/\phi)^{-1};}
& m=2, \\
{\displaystyle E_c^m(\phi_c/\phi)^{2-m};} \hskip 1cm
& m\geq 3,
\end{array} \right.
                                                           \label{new}
\end{equation}
As is shown in  Appendix A, this structure of the moments is not an
artefact of the simple model, but may be rigorously derived from a general
$N\times N$ random Hamiltonian, Eq.\ (\ref{ham}). Based on this, one may
suggest  the universality of the large curvature,
$|\partial^2\epsilon_n/\partial \phi^2|\gg E_c$, behavior   of the
distribution, Eq.\ (\ref{new1}).
In contrast, the small curvature part,
$|\partial^2\epsilon_n/\partial \phi^2|\ll E_c$, may be a specific
property of our toy model.

Equation (\ref{new}) definitely contradicts  the frequently used expression,
Eq.\
(\ref{second}), when evaluating the average over the disorder as arithmetical
mean. Indeed, according to Eq.\ (\ref{new}) the second and higher
moments of the curvature diverge at zero flux (or, in other words, in the GOE
ensemble). This is a consequence of the absence of the exponential tail in
the distribution function at exactly zero flux. In this case the behavior
$P_0(c)\sim c^{-3}$ continues up to infinity. As was already mentioned, such a
behavior is a common feature of GOE and not a result of an oversimplified
model (see Appendix A). Thus when evaluating the
Thouless energy, Eq.\ (\ref{second}), one should either use a geometrical mean,
as already pointed out by Thouless \cite{Thouless77}, who assumed a simple
Lorentzian distribution of the curvatures, or use another measure
of the sensivity of the spectrum to variations of the boundary conditions:
$$E_c\sim\langle\left|\left.\frac{\partial^2\epsilon_n}{\partial \phi^2}
\right|_{\phi=0}\right|\rangle.$$

The same divergence affects dramatically the current--current correlation
function. Namely, it will be shown that for $\phi,\phi ' \ll\phi_c $,
one has
$$
\langle\frac{\partial\epsilon_n(\phi)}{\partial \phi}
\frac{\partial\epsilon_n(\phi ')}{\partial \phi '} \rangle\sim
-E_c^2\phi\, \phi '\ln\left(\frac{\phi+\phi '}{\phi_c}\right).
$$
Differentiating this result with respect to $\phi$ and $\phi '$ and then
putting $\phi ' = \phi$, one returns back to the just quoted second moment
of the curvature. This  expression is absolutely unexpected from the point of
view of  perturbation theory. The latter assumes rather
$E_c^2\phi\, \phi '$ (without logarithm) for the above defined correlation
function. It would be definitely interesting to see if such a behavior exists
in
the framework of supersymmetric calculations. In fact,  these calculations have
already been done in Ref.\ \cite{Simons93}, but, as far as we know, only in
GUE,
where we do not expect anything unusual.

Curvature distributions  have  already been investigated in a number of works
\cite{Gaspard90,Takami92,Delande93},
however, in a very different context. The
considered Hamiltonians had the structure $H(\lambda)=H_1+\lambda H_2$,
where $H(\lambda)$  belongs to the {\em same} universality class in the whole
range of $\lambda$ \cite{Delande93}.
The curvature is defined as the second derivative of
energy with respect to the parameter $\lambda$. Subsequently, curvature
distributions for three pure symmetry classes were studied. Since
the above defined parameter $\lambda$ does not break the time--reversal
symmetry, it  can not play the role of a magnetic flux. Thus, it has no
meaning, for example, to look for a current (defined as a derivative
with respect to $\lambda$)
distribution. For reasons, which become clear in  Appendix A, our
results for $\alpha=0$ coincide exactly with the GOE results of Refs.\
\cite{Gaspard90,Takami92,Delande93}.
We recover the $c^{-3}$ tail of the curvature distribution, first discovered
in Ref.\ \cite{Gaspard90}, and Eq.\ (\ref{goec}) of the present work may be
found in Ref.\ \cite{Delande93}. However, for any $\alpha\neq 0$ our
conclusions are very different from those of Refs.\
\cite{Gaspard90,Takami92,Delande93}.
For example, in GUE ($\alpha=1$) we have find a Gaussian tail of
the curvature (with respect to flux) distribution. At the same time the
distribution function of the curvatures, defined with respect to the parameter
$\lambda$, decays only as the fourth power (see Appendix A and Ref.\
\cite{Gaspard90}).

The present article has the following structure. In section \ref{s2} we
specify the $2\times 2$ model based on Eq.\ (\ref{ham}) and re--derive
the known
results concerning the energy spacing distribution. In sections \ref{s3} and
\ref{s4}  single level current and curvature DF's are derived.
Finally
in section \ref{s5} we discuss the possible implications of the simple model
to real physical systems. In Appendix A  the moments of the
curvature for a general $N\times N$ random matrix model are considered.
A summary of the results for pure ensembles (GOE and
GUE) is given in  Appendix B.


\section{The model and energy spacing distribution}
\label{s2}

Consider a model based on $H^{(2)}(\alpha)$, as defined by Eq.\ (\ref{ham})
\begin{equation}
H^{(2)}(\alpha)=\left(\begin{array}{cc}
   x_1+x_2           \hskip 1cm  & x_3+i\alpha x_4 \\
x_3-i\alpha x_4      \hskip 1cm  &    x_1-x_2
\end{array}\right),
                                                         \label{matrix}
\end{equation}
where $x_j (j=1\ldots 4)$ are real random variables, with a Gaussian
distribution law
\begin{equation}
P(x_j)=\frac{1}{v\sqrt{2\pi}}
\exp\left(-\frac{x_j^2}{2 v^2}\right)\,.
                                                          \label{gauss}
\end{equation}
The variance of the distribution $v^2$ and the time--reversal symmetry
breaking parameter $\alpha$ are the two free parameters of the model.
In the end
 they should be related to physical observables such as mean level
spacing and magnetic flux. Let us, however, postpone this discussion until
section \ref{s5}. The spectrum of the Hamiltonian, Eq.\ (\ref{matrix}), is
given by
\begin{equation}
\epsilon_{\pm}=x_1\pm\left(x_2^2+x_3^2+\alpha^2 x_4^2\right)^{1/2},
                                                          \label{spectrum}
\end{equation}
and the energy spacing $\epsilon(\alpha,{\bf x})$ by
\mbox{$\epsilon(\alpha,{\bf x})\equiv\epsilon_+ -\epsilon_-
=
2\left(x_2^2+x_3^2+\alpha^2 x_4^2\right)^{1/2}$}. Let us consider an energy
spacing DF
\begin{equation}
P_{\alpha}(\epsilon)=\int\delta(\epsilon-\epsilon(\alpha,{\bf x}))
\prod_{j=1}^4 P(x_j)d x_j.
                                                          \label{def}
\end{equation}
The calculation of the integrals is straightforward, finally one obtains
\cite{French88}
\begin{equation}
P_{\alpha}(\epsilon)=\frac{\epsilon}{4v^2\sqrt{1-\alpha^2}}
\exp\left(-\frac{\epsilon^2}{8v^2}\right)
\mbox{erf}\left(\sqrt{\frac{1-\alpha^2}{2\alpha^2}}\frac{\epsilon}{2v}\right).
                                                          \label{adistr}
\end{equation}
In two limiting cases one  returns again to the familiar distributions:
for $\alpha=0$
\begin{equation}
P_0(\epsilon)=\frac{\epsilon}{4 v^2}
\exp\left(-\frac{\epsilon^2}{8v^2}\right); \hskip 3.5cm \mbox{(GOE)},
                                                          \label{goee}
\end{equation}
and for $\alpha=1$
\begin{equation}
P_1(\epsilon)=\frac{\epsilon^2}{4 \sqrt{2\pi}v^3}
\exp\left(-\frac{\epsilon^2}{8v^2}\right);  \hskip 2.5cm \mbox{(GUE)},
                                                          \label{guee}
\end{equation}
It is a well--known fact that in the case of GOE,
$P_0(\epsilon)\propto\epsilon$
for $\epsilon\ll v$, whereas in the absence of  time--reversal symmetry (GUE)
the level repulsion is stronger: \mbox{$P_1(\epsilon)\propto\epsilon^2$};
$(\epsilon\ll v)$. In the intermediate region $0<\alpha<1$ one has
\begin{equation}
P_{\alpha}(c)\approx \frac{1}{v}\exp\left(-\frac{\epsilon^2}{8v^2}\right)
\left\{ \begin{array}{ll}
{\displaystyle
\left(\frac{\epsilon}{2v}\right)^2\frac{1}{\sqrt{2\pi}\alpha};}
& {\displaystyle \epsilon\ll \frac{v\alpha}{\sqrt{1-\alpha^2}},}  \\
{\displaystyle \left(\frac{\epsilon}{2v}\right)\:
\frac{1}{2\sqrt{1-\alpha^2}};} \hskip .5cm
& {\displaystyle \epsilon\gg \frac{v\alpha}{\sqrt{1-\alpha^2}},}
\end{array} \right.
                                                         \label{adistr1}
\end{equation}
The level repulsion is quadratic for small energy intervals, and becomes
linear for larger ones. The moments of the distribution are given by
\begin{eqnarray}
\langle\epsilon^m\rangle_{\alpha} = v^m \left\{ \begin{array}{ll}
{\displaystyle \frac{(-1)^n}{\sqrt{\pi}} \frac{2^{3n-1/2}}{\sqrt{1-\alpha^2}}
\, \frac{\partial^n}{\partial t^n} \left[ t^{-1/2}\arctan\sqrt{\frac
{1-\alpha^2}{\alpha^2t}}\right]\left|_{t=1}\right.;}
& m=2n-1, \\
{\displaystyle (-1)^n 2^{3n+1}\, \frac{\partial^n}{\partial t^n}\left[
t^{-1}(\alpha^2t+1-\alpha^2)^{-1/2}\right]\left|_{t=1}\right.;}
& m=2n.
\end{array} \right.
                                                          \label{mome}
\end{eqnarray}
In particular for the first moment one has \cite{French88}
\begin{equation}
\langle\epsilon\rangle_{\alpha}=v\sqrt{\frac{8}{\pi}}
\left[\alpha+ (1-\alpha^2)^{-1/2}\arctan\sqrt{\frac
{1-\alpha^2}{\alpha^2}}\,\right].
                                                          \label{ave}
\end{equation}
This is a smooth monotonous function of $\alpha$, which varies from
$\langle\epsilon\rangle_0=v\sqrt{\frac{8}{\pi}}\, \frac{\pi}{2}$ up to
$\langle\epsilon\rangle_1=v\sqrt{\frac{8}{\pi}}\, 2$. The fact that it is
almost
constant will be useful  in section \ref{s5}, where we shall try to
give a physical interpretation of the results.

\section{Single level current distribution function}
\label{s3}

Define now the single level currents in a $2\times 2$ model as
\begin{equation}
i_{\pm}(\alpha,{\bf x})\equiv
\frac{\partial \epsilon_{\pm}}{\partial \alpha}=
\pm\frac{2\alpha x_4^2}{\epsilon(\alpha,{\bf x})}\, .
                                                          \label{current}
\end{equation}
We will look for a DF, $P_{\alpha}(i)$ of
$i\equiv i_+\geq 0$. Obviously the distribution of $i_-$ is the same (up to the
minus sign of the argument).
If one is interested in the DF of $\tilde{i}$, which may have either sign, one
simply has
$\tilde{P}_{\alpha}(\tilde{i})=P_{\alpha}(|\tilde{i}|)/2$
(the coefficient $1/2$
takes care of the correct normalization). To evaluate $P_{\alpha}(i)$ let us
 first calculate the joint (current and energy space) distribution,
$P_{\alpha}(i,\epsilon)$. Besides  technical advantages, this way of
calculation provides some additional information. Namely, one will be able to
identify those energy spacings $\epsilon$, that are mainly
responsible for a given current, $i$. The joint DF is defined as
\begin{equation}
P_{\alpha}(i,\epsilon)=\int_{-\infty}^\infty\delta(i-i(\alpha,{\bf x}))
\delta(\epsilon-\epsilon(\alpha,{\bf x})) \prod_{j=1}^4 P(x_j)d x_j.
                                                          \label{defie}
\end{equation}
After some calculations one gets
\begin{equation}
P_{\alpha}(i,\epsilon)=\theta(\epsilon-2\alpha i)
\frac{1}{2}\frac{\pi}{(v\sqrt{2\pi})^3}
\frac{\epsilon^{3/2}}{\sqrt{2\alpha i}}
\exp\left(-\frac{\epsilon^2 \alpha^2+2\epsilon\alpha (1-\alpha^2) i}
{8v^2 \alpha^2} \right)\,.
                                                          \label{distrie}
\end{equation}
Here $\theta(x)$ is a usual step function (remember that both $i,\epsilon\geq
0$). To get the current distribution, $P_{\alpha}(i)$, one should, according
to Eq.\ (\ref{defie}), integrate the last expression over $\epsilon$.
\begin{eqnarray}
P_{\alpha}(i)=\frac{1}{2}\frac{\pi}{(v\sqrt{2\pi})^3}(2\alpha i)^2
\int_0^{\infty} (1+t)^{3/2}
\exp\left(-\frac{i^2}{2v^2}(t^2 \alpha^2+ t(1+\alpha^2)+1)\right) dt,
                                                               \label{distri}
\end{eqnarray}
where a variable $t$ was introduced as $\epsilon=2\alpha i(t+1)$.
The last integral is not known in special functions, except in the two limiting
cases $\alpha=0$ and $\alpha=1$ (see below). However one can work out it's
asymptotic behavior in various regions
\begin{equation}
P_{\alpha}(i)\approx\frac{1}{v}\left\{ \begin{array}{lll}
{\displaystyle \frac{\Gamma(1/4)}{2^{9/4}\sqrt{\pi}}\,
\sqrt{\frac{v}{i\alpha}};}
& i\ll v\alpha; &  (\epsilon\approx v) \\
{\displaystyle 3 \alpha^2 \left(\frac{v}{i}\right)^3;}
& v\alpha\ll i \ll v; \hskip.5cm  & (\epsilon\approx \alpha i) \\
{\displaystyle \sqrt{\frac{2}{\pi}}\, \frac{\alpha^2}{1+\alpha^2}
\exp\left(-\frac{i^2}{2 v^2}\right);} \hskip .5cm
& v\ll i; &(\epsilon\approx \alpha v)\,.
\end{array} \right.
                                                         \label{idistr}
\end{equation}
The values of energy spacings $\epsilon$, which provide the main
contribution in each case are designated in brackets.
Equation (\ref{idistr}) shows that the
single level current DF has an integrable square root singularity at small
currents. Realizations with  energy spacings of the order of an average one are
mostly responsible for this singularity. In the intermediate region, which
exists only if $\alpha\ll 1$, the current DF decreases as $i^{-3}$. Finally,
for large currents the distribution has a Gaussian tail. For $\alpha\ll 1$
the last two regions arise due to realizations with extremely small energy
spacings.

For the two pure cases one can calculate the distributions analytically: for
$\alpha=0$
\begin{equation}
P_0(i)=\frac{1}{v}2\delta\left(\frac{i}{v}\right),
\hskip 5.5cm \mbox{(GOE)}
                                                          \label{goei}
\end{equation}
which is evident without any calculations, and for $\alpha=1$
\begin{equation}
P_1(i)=\frac{1}{v}\frac{1}{\sqrt{2\pi}}\sqrt{\frac{\sqrt{2}v}{i}}\,
\Gamma\left(\frac{5}{4},\left(\frac{i}{\sqrt{2}v}\right)^2\right),
\hskip 1.5cm \mbox{(GUE)}
                                                          \label{guei}
\end{equation}
where $\Gamma(a,x)$ is an incomplete gamma function. The asymptotic behavior
of $P_1(i)$ is given by the first and the third lines of Eq.\ (\ref{idistr})
(with $\alpha=1$).
The moments of the current distribution are given by
\begin{equation}
\langle i^m\rangle_{\alpha\ll 1}\approx v^m\left\{ \begin{array}{ll}
{\displaystyle \sqrt{\pi/2}\, \alpha\, ;}
& m=1, \\
{\displaystyle -3 \alpha^2 \ln\alpha\, ;}
& m=2, \\
{\displaystyle \frac{2^{m/2+1}}{\sqrt{\pi}}\alpha^2
\Gamma\left(\frac{m+3}{2}\right)\frac{1}{m-2}\, ;} \hskip .5cm
& m\geq 3,
\end{array} \right.
                                                         \label{momi}
\end{equation}
for $\alpha\ll 1$ in  leading order in $\alpha$, and
\begin{equation}
\langle i^m\rangle_{\alpha\approx 1} \approx v^m \frac{2^{m/2}}{\sqrt{\pi}}
\Gamma\left(\frac{m+3}{2}\right)\frac{1}{m+1/2}\, ,
                                                          \label{momi1}
\end{equation}
for $\alpha\approx 1$ in  leading order in $(1-\alpha^2)$.
Let us mention particularly that the second moment of a single level current
is given by $\langle i^2\rangle_{\alpha\ll 1}\approx -3v^2\alpha^2\ln\alpha$.
This can be
seen directly from  Eq.\ (\ref{idistr}). The leading term comes from the
intermediate region of the currents ($v\alpha\ll i\ll v$). The main
contribution
to $\langle i^2\rangle_{\alpha\ll 1}$ arises from realizations with very
small spacings $\epsilon\approx \alpha^2v\ll \langle \epsilon\rangle$. Starting
from the third one, all the moments are determined by a Gaussian tail of the
distribution ($v\ll i$), whereas the average current
$\langle i\rangle_{\alpha\ll 1}$ comes from the opposite region $i\ll v$.

\section{Curvature distribution function}
\label{s4}
Following the same scheme as in the previous section,  consider now the single
level curvature DF.
\begin{equation}
c_{\pm}(\alpha,{\bf x})\equiv
\frac{\partial^2 \epsilon_{\pm}}{\partial \alpha^2}=
\pm\left(\frac{2 x_4^2}{\epsilon(\alpha,{\bf x})}-
\frac{8 \alpha^2 x_4^4}{\epsilon^3(\alpha,{\bf x})}\right).
                                                          \label{curv}
\end{equation}
Again we consider a distribution of a positive defined quantity $c\equiv c_+
\geq 0$ (it is indeed positive as $\epsilon\geq 2\alpha x_4$, according to
Eq.\ (\ref{spectrum})). If one wants to include also $c_-$, one should again
make the distribution symmetrical with a proper normalization. The joint
(curvature,
energy spacing) DF, $P_{\alpha}(c,\epsilon)$, may be found in elementary
functions in the closed form
\begin{eqnarray}
P_{\alpha}&&(c,\epsilon)=\theta(\epsilon-8\alpha^2 c)
\frac{1}{4}\frac{\pi}{(v\sqrt{2\pi})^3}
\frac{\epsilon^{3/2}}{\sqrt{c}}
\exp\left(-\frac{\epsilon^2}{8 v^2}\frac{1+\alpha^2}{2 \alpha^2}\right)\times
\nonumber \\
&&B^{-1}\left[
\sqrt{1+B}\exp\left(\frac{\epsilon^2}{8 v^2}\frac{1-\alpha^2}{2 \alpha^2}
B \right) +
\sqrt{1-B}\exp\left(-\frac{\epsilon^2}{8 v^2}\frac{1-\alpha^2}{2 \alpha^2}
B \right) \right],
                                                          \label{distrce}
\end{eqnarray}
where $B=\sqrt{1-8\alpha^2 c/\epsilon}$. One now integrates the last expression
over $\epsilon$ and obtains the following asymptotics for the curvature DF:
\begin{equation}
P_{\alpha}(c)\approx \frac{1}{v} \left\{ \begin{array}{lll}
{\displaystyle \frac{\Gamma(1/4)}{2^{9/4}\sqrt{\pi}}\, \sqrt{\frac{v}{c}}\, ;}
& c\ll v, &  (\epsilon\approx v) \\
{\displaystyle 3 \left(\frac{v}{c}\right)^3;}
& {\displaystyle v\ll c \ll\frac{v}{\alpha},}& (\epsilon\approx \alpha^2 c) \\
{\displaystyle  \frac{2^{5/2}\alpha^4}{\sqrt{1+\alpha^2}} \frac{c}{v}
\exp\left(-\frac{\alpha^2 c^2}{2 v^2}g_{\alpha}\right)}
\left\{ \begin{array}{l}
1;\\
2;
\end{array} \right.
\hskip .3cm &
\begin{array}{l}
{\displaystyle \frac{v}{\alpha}\ll c\ll \frac{v}{\alpha(1-\alpha^2)},} \\
{\displaystyle \frac{v}{\alpha(1-\alpha^2)} \ll c,}
\end{array}
& (\epsilon\approx \alpha v)
\end{array} \right.
                                                         \label{cdistr}
\end{equation}
where $g_{\alpha}=(7\alpha^4+18\alpha^2+7)/(1+\alpha^2)$. The first two lines
in Eq.\ (\ref{cdistr}) look very similar to those of the current
distribution (cf. Eq.\ (\ref{idistr})). The reason is that for $c\ll
v/\alpha$ the last term in the expression for the curvature,
Eq.\ (\ref{curv}), may be omitted, thus one has a
trivial relationship between current and curvature, $i=\alpha c$. The large
curvature ($c\gg v/\alpha$) tail of the distribution is affected by the last
term of Eq.\ (\ref{curv}), resulting in a complicated form of the tail of the
DF, Eq.\ (\ref{cdistr}). For $\alpha=0$ (GOE) the tail disappears completely,
in this case one can get an exact result for the curvature distribution
(see also Ref.\ \cite{Delande93})
\begin{equation}
P_0(c)=\frac{3}{8v}\sqrt{\frac{2v}{c}} \exp\left(\frac{1}{4}
\left(\frac{c}{2v}\right)^2\right) D_{-5/2}\left(\frac{c}{2v}\right),
\hskip 2cm \mbox{(GOE)}
                                                          \label{goec}
\end{equation}
where $D_a(x)$ is a Whittaker parabolic cylinder function. The asymptotic
behavior of this DF is given by the first two lines of the general expression,
Eq.\ (\ref{cdistr}). Due to the absence of an exponential tail in a GOE
curvature distribution, Eq.\ (\ref{goec}),
all the moments of it, starting from the
second one, diverge. Indeed, a direct evaluation of the moments in leading
order in $\alpha$ results in
\begin{equation}
\langle c^m\rangle_{\alpha\ll 1}\approx v^m\left\{ \begin{array}{ll}
{\displaystyle \sqrt{\pi/2};}
& m=1, \\
{\displaystyle -3 \ln\alpha;}
& m=2, \\
{\displaystyle \frac{2^{m/2+1}}{\sqrt{\pi}}\alpha^{2-m}
\Gamma\left(\frac{m+3}{2}\right)
\sum_{p=0}^m\frac{(-1)^p\left(\stackrel{\textstyle m}{\textstyle p}\right)}
{m-2+2p};} \hskip .5cm
& m\geq 3,
\end{array} \right.
                                                         \label{momc}
\end{equation}
for $\alpha\ll 1$ (cf. with Eq.\ (\ref{momi})), and
\begin{equation}
\langle c^m\rangle_{\alpha\approx 1}\approx v^m\frac{2^{m/2}}{\sqrt{\pi}}
\Gamma\left(\frac{m+3}{2}\right)
\sum_{p=0}^m\frac{(-1)^p\left(\stackrel{\textstyle m}{\textstyle p}\right)}
{m+1/2+p},
                                                          \label{momc1}
\end{equation}
for $\alpha\approx 1$ (cf. with Eq.\ (\ref{momi1}));
$\left(\stackrel{\textstyle m}{\textstyle p}\right)$ is a binomial
coefficient.
As in the case of the moments of the current, all the moments of the curvature,
starting from the second one arise from  realizations with small energy
spacings $\epsilon\ll\langle\epsilon\rangle$ (if $\alpha\ll 1$). At $\alpha=0$
only the first moment exists, whereas all  higher moments diverge. This
unexpected fact will be discussed in more detail in the next section.

\section{Discussion of the results}
\label{s5}

As already mentioned in section \ref{s1}, the Wigner surmise
obtained for $2\times 2$ matrices, works extremely well also for a large $N$.
To establish this connection one should relate the phenomenological parameter
$v^2$ -- the variance of the distribution -- to an average level spacing
$\Delta$. One
simply demands that
$$\langle\epsilon\rangle_{\alpha}=\Delta.$$
Strictly speaking the average spacing $\langle\epsilon\rangle_{\alpha}$ is a
function of $\alpha$ (magnetic flux), although the mean level spacing
$\Delta$ is presumably a constant, independent of external parameters.
However, as we noticed after Eq.\ (\ref{ave}), the dependence on $\alpha$ is
very week (especially for small $\alpha$). Using this fact we shall
disregard its $\alpha$--dependence and just admit
\begin{equation}
\langle\epsilon\rangle_0=\Delta=v\sqrt{2\pi},
                                                          \label{id1}
\end{equation}
where we have used Eq.\ (\ref{ave}). Being honest, one should re--identify
the parameters
for each value of $\alpha$ separately. This procedure, although trivial, is
not transparent enough for our illustrative purposes.

Having an energy spacing DF as an example, one may hope that the single level
current
and curvature DF's, derived for a $2\times 2$ system, may be suitable for
larger systems as well. To support the last statement let us put forward the
following arguments. As we have seen in previous sections, all the moments of
a single level current and curvature DF's, starting from the second one,
arise mainly from realizations of a random Hamiltonian with  very small
gaps. For these realizations the $2\times 2$ ansatz is supposed to be
essentially correct, because for the close pair of levels only their mutual
interaction appears to be important. In  Appendix A we prove that the
small flux behavior of the moments is indeed observed in the general  $N\times
N$ model as well.
The first moment, however, is
determined by realizations with an energy gap of the order of the average
one. In this case  the $2\times 2$ scheme need not  be precise.
Thus one  should
not trust the value of the first moment, but rather connect it
phenomenologically with  the microscopic characteristics of a system. Following
Thouless, one may relate a typical second derivative (not \mbox{r.m.s !})
with respect to flux at zero flux to a correlation energy
\begin{equation}
E_c=\langle\left|\left.\frac{\partial^2\epsilon_{\pm}}{\partial \phi^2}
\right|_{\phi=0}\right|\rangle,
                                                          \label{first}
\end{equation}
cf. with Eq.\ (\ref{second}). On the other hand, we had (see Eq's.\
(\ref{curv}), (\ref{momc}))
$$\langle\left|\left.\frac{\partial^2\epsilon_{\pm}}{\partial \alpha^2}
\right|_{\alpha=0}\right|\rangle=v\sqrt{\frac{\pi}{2}}.$$
Using the definition of a mean level spacing, Eq.\ (\ref{id1}), one obtains
\begin{equation}
\alpha=\sqrt{\frac{2 E_c}{\Delta}}\phi,
                                                            \label{id2}
\end{equation}
This should be compared with the conjecture of Dupuis and Montambaux
\cite{Montambaux91}, Eq.\ (\ref{conj}) (with $N=2$). As one sees, the
agreement is extremely good, the slight discrepancy may be attributed to the
fact that Eq.\ (\ref{conj}) was obtained for the large $N$ limit. We
conjecture thus, that with the identifications, Eq's.\ (\ref{id1}),
(\ref{id2}), the tails of the distributions obtained for a $2\times 2$ model
are applicable
for larger systems as well. Let us discuss the further consequences of this
rather strong assumption.

First of all one notices that the $\alpha=1$ (GUE) case corresponds to the
value of
a flux $\phi_c=\sqrt{\Delta/(2 E_c)}$. This value is well--known
as a correlation flux. Up to this flux a typical level may change
parabolically, without crossing  other levels. At $\phi=\phi_c$ the first
avoiding crossing event usually happens, and the simple $2\times 2$ scheme
obviously breaks down. It was demonstrated numerically \cite{Montambaux91},
that at $\phi\approx \phi_c$ the crossover to GUE is indeed practically
completed. This shows that the applicability of  a $2\times 2$ model for
$0\leq\phi\leq\phi_c$ ($0\leq\alpha\leq 1$) is quite reasonable, as well as
the identification of the $\phi=\phi_c$ point with GUE.

Consider now the second moment of the single level current in GUE (or, the
same, at $\phi=\phi_c$). Using Eq.\ (\ref{momi1}), one obtains
$(\langle i^2 \rangle_{\alpha=1})^{1/2}=v\sqrt{3/5}$, or in physical parameters
(using
Eq's.\ (\ref{id1}), (\ref{id2}))
$$\left[\langle\left(\left.\frac{\partial\epsilon_{\pm}}{\partial \phi}
\right|_{\phi=\phi_c}\right)^2\rangle\right]^{1/2}=
\sqrt{\frac{3}{5\pi}}\, \sqrt{\Delta E_c}.$$
This result is also well-known from  perturbation theory  (up to
the numerical coefficient) \cite{Gefen91,Oppen91}.

Being thus  convinced that the obtained results lead to reasonable
predictions for real physical systems, let us  discuss the most
surprising feature of the considered DF's: At zero flux all
the moments of the curvature, starting from the second one, {\em diverge}.
Thus, when calculating the correlation energy from the curvatures, one has to
use another measure for their typical value than just the root mean square,
like Eq.\ (\ref{first}) or the geometrical mean proposed by Thouless
\cite{Thouless77}.

The consequences of the discussed divergence are, however, deeper than just
the neccessity of a more careful definition of the correlation energy. One also
should reconsider the
universal relationship between dissipative and correlation conductances,
derived
by Akkermans and Montambaux \cite{Akkermans91}. Mathematically this relation
was expressed as \cite{Akkermans91}
$$\overline{\langle\left[\frac{\partial\epsilon_n}{\partial \phi}
\right]^2\rangle}=a\Delta
\langle\left[\left. \frac{\partial^2\epsilon_n}{\partial \phi^2}
\right|_{\phi=0}\right]^2\rangle^{1/2},$$
where bar denotes integration with respect to flux, and $a$ is a
universal numerical factor.
According to the present results this relation can not hold, when the typical
curvature is calculated as an arithmetical mean. Indeed, using
Eq.\ (\ref{distri}) and Eq's.\ (\ref{id1}),
(\ref{id2}),  one obtains for the l.h.s. of the last expression $\sqrt{\Delta
E_c}$ (up to a coefficient of the order of unity), whereas the r.h.s. diverges.

To understand the reason for this phenomena let us consider the two point
current--current correlation function
\begin{equation}
{\bf C}(\alpha,\alpha ')\equiv \langle i(\alpha)i(\alpha ')\rangle.
                                                          \label{corr}
\end{equation}
A very similar object was recently considered in Ref.\ \cite{Altshuler93}.
One can explicitly perform the averaging in a  $2\times 2$ model by integrating
over
$d{\bf x}$ with the corresponding weight, precisely as one did in the previous
sections. The general answer is cumbersome, but one needs only the  behavior
for  small flux. In this case one easily gets
\begin{equation}
{\bf C}(\alpha,\alpha')\approx -3v^2\alpha\alpha '\ln(\alpha+\alpha ')\, ;
\hskip 2cm   \alpha , \alpha '\ll 1.
                                                          \label{corr1}
\end{equation}
Putting here $\alpha =\alpha '$, one  returns again to the expression for the
second moment of the current (the second line in Eq.\ (\ref{momi})). On the
other
hand, differentiating Eq.\ (\ref{corr1})  with respect to $\alpha$ and
$\alpha '$ and then putting  $\alpha =\alpha '$, one recognizes the second
line of Eq.\ (\ref{momc}). In physical parameters Eq.\ (\ref{corr1}) may be
rewritten as
\begin{equation}
{\bf C}(\phi,\phi ')\equiv
\langle\frac{\partial\epsilon_{\pm}(\phi)}{\partial \phi}
\frac{\partial\epsilon_{\pm}(\phi ')}{\partial \phi '} \rangle\approx
-6\pi^{-1} E_c^2\phi\, \phi '\ln\left(\frac{\phi+\phi '}{\phi_c}\right);
\hskip 1cm \phi,\phi '\ll \phi_c.
                                                          \label{corr3}
\end{equation}
This should be compared with the corresponding result of the perturbative
calculations
\begin{equation}
{\bf \tilde{C}}(\phi,\phi ')\approx
12\pi^{-2} E_c^2\phi\, \phi '\left(\frac{\Delta}{\gamma}\right)^2;
\hskip 4cm \phi,\phi '\ll \phi_c,
                                                          \label{corr4}
\end{equation}
where $\gamma$ is a cut off in perturbation theory, which is usually supposed
to be of the order of $\Delta$ \cite{Montambaux91}.  The discrepancy between
the two results is rather dramatic. Whereas Eq.\ (\ref{corr4}) leads to a
finite
second moment of the curvature ($\approx E_c^2$), Eq.\ (\ref{corr3}) results in
a
divergent second moment. Let us also point out that the perturbative result
obviously may be expressed in a form
${\bf \tilde{C}(\phi,\phi ')}=f(\phi+\phi ')-f(\phi-\phi ')$, which may be
traced back to
Diffuson and Cooperon channels in the diagrammatic expansion.
The Eq.\ (\ref{corr3}) does not allow such a
decomposition. This might be a point where the present scheme contradicts the
derivation of  Ref.\ \cite{Akkermans91}. Indeed, it was
assumed explicitly \cite{Akkermans91}, that the Diffuson--Cooperon
decomposition (which is certainly correct for a large  flux $\phi\gg\phi_c$) is
also valid in the vicinity of zero flux. According to the present
consideration this is not the case.

It is not clear at the moment whether the discussed divergence has a real
physical meaning,
but if so, it  might cause difficulties in numerical calculations
of correlation functions in GOE. We conclude that further analytical
(both RMT--like and supersymmetric) calculations and
numerical work are necessary to clarify this unexpectedly controversial issue.

\section{Acknowledgements}
\label{s10}
We are very grateful to Yuval Gefen, Gilles Montambaux, Erik Akkermans and
Uzy Smilansky for numerous and helpful discussions. We want to acknowlege the
hospitality of the Institute for Scientific Interchange (ISI), Torino, where
this
work was completed. One of us (A.K.) was supported by the German--Israel
Foundation (GIF) and the U.S.--Israel Binational Science Foundation (BSF).

\appendix{Moments of the curvature in a $N\times N$ model}
\label{s8}
Let us show now that the flux dependence of the moments of
curvature, derived for a $2\times 2$ model, Eq.\ (\ref{momc}), and summarized
in Eq.\ (\ref{new}), may be obtained from a general $N\times  N$ model.
Consider  an $N\times N$ random matrix  Hamiltonian, given by
Eq.\ (\ref{ham}).  Without loss of generality one may assume
that its spectrum  is not degenerate. Then, using second order
perturbation theory, one obtains the following {\em exact} relationship
\begin{equation}
\frac{\partial^2\epsilon_n(\alpha)}{\partial\alpha^2}=
2\sum_{k\neq n}^N\frac{\left|<\alpha,k|H_A^{(N)}|n,\alpha>\right|^2}
{\epsilon_n(\alpha)-\epsilon_k(\alpha)},
                                                        \label{pert}
\end{equation}
where $\epsilon_k(\alpha)$ and $|k,\alpha>$ are eigenvalues and
eigenfunctions of the full  Hamiltonian, $H^{(N)}(\alpha)$.
As it is well-known from RMT
\cite{Mehta91}, statistics of eigenvalues and statistics of
eigenfunctions are completely independent of each other.

Let us first consider the case of exactly zero flux ($\alpha=0$, GOE).
In this case the energies in the denominator on the r.h.s of Eq.\ (\ref{pert})
are
eigenvalues of $H_S^{(N)}$, whereas in the numerator one has matrix elements of
$H_A^{(N)}$. Hence,  matrix elements,
$\left|<0,k|H_A^{(N)}|n,0>\right|^2$, and eigenvalues,
$\epsilon_k\equiv \epsilon_k(0)$, may be
considered as independent random variables. The statistic of the eigenvalues is
given by a Wigner--Dyson (GOE) distribution \cite{Mehta91}
\begin{equation}
P_N(\epsilon_1,\ldots,\epsilon_N)=
\mbox{const}\times\exp\left(-\frac{1}{2}\sum_{k=1}^N\epsilon_k^2\right)
\prod_{1\leq k \leq n \leq N}|\epsilon_n-\epsilon_k|,
                                                        \label{WD}
\end{equation}
whereas an exact form of the matrix element
distribution  is not important for our  purposes. One is
now in a position to consider the moments of the random variable
\mbox{$c=
\left|\partial^2\epsilon_n(\alpha)/\partial\alpha^2|_{\alpha=0}\right|$}.
Doing this, one will be interested
only in the manner of divergence, omitting all the prefactors as well as less
divergent terms. Raising Eq.\ (\ref{pert}) to the m$^{th}$ power and
averaging, one obtains
\begin{equation}
\langle c^m \rangle\sim N^{-1}\int\!\!\!\int
\frac{R_2(\epsilon_1,\epsilon_2)}{|\epsilon_1-\epsilon_2|^m}
d\epsilon_1 d\epsilon_2\, +\, \ldots \, ,
                                                        \label{nm}
\end{equation}
where averaging over matrix elements, $\left|<0,k|H_A^{(N)}|n,0>\right|^2$,
leads to some omitted constant prefactor, and  ``$\ldots$'' denotes less
divergent terms, arising from the non--diagonal contributions, like
$$\int\!\!\!\int\!\!\!\int
\frac{R_3(\epsilon_1,\epsilon_2,\epsilon_3)}
{|\epsilon_1-\epsilon_2|^{m-1}|\epsilon_2-\epsilon_3|}
d\epsilon_1 d\epsilon_2 d\epsilon_3\, ,
$$
{\em etc}. Here $R_i(\epsilon_1,\ldots,\epsilon_i)$ is an $i$--point
correlation function \cite{Mehta91}, for example
\begin{equation}
R_2(\epsilon_1,\epsilon_2)\equiv\frac{1}{N(N-1)}
\int_{-\infty}^{\infty}\ldots \int_{-\infty}^{\infty}
P_N(\epsilon_1,\ldots,\epsilon_N) d\epsilon_3\ldots d\epsilon_N.
                                                        \label{r2}
\end{equation}
In the limit of large $N$, the correlation functions $R_i$  depend only on
differences
of the eigenvalues. Then the integral over $\epsilon_1+\epsilon_2$
in Eq.\ (\ref{nm}) leads to some constant of the order of $N$, and one finally
obtains
\begin{equation}
\langle c^m \rangle\sim \int_{0}^{\infty}
\frac{R_2(\epsilon)}{\epsilon^m} d\epsilon ,
                                                        \label{lam}
\end{equation}
where $\epsilon=|\epsilon_1-\epsilon_2|$. Taking into account
the  well-known  GOE result, $R_2(\epsilon)\sim\epsilon$ ($\epsilon\ll
\Delta$), one notices
that all the moments of the curvature, starting from the
second one, diverge. This is in exact agreement with the result
for a $2\times 2$ model, but now one has demonstrated the validity of this
statement for a general $N\times N$ model.

The above developed scheme is applicable without any
changes for the case where $\alpha$ is not a time--reversal symmetry breaking
parameter, hence $H^{(N)}(\alpha)$ belongs to the same universality class in
the whole range of $\alpha$ \cite{Gaspard90,Delande93}. Indeed, in this case
the matrix elements of the perturbation (numerator on r.h.s of Eq.\
(\ref{pert}))
and the
eigenenergies (denominator) can always be considered as independent random
variables. One can, for example, repeat exactly the same arguments for GUE
with a single modification: $R_2(\epsilon)\sim\epsilon^2$. Then one obtains
the divergence of all the moments of the curvature starting from the {\em
third}
one. This perfectly agrees with the result for the tail of the curvature
distribution, proved in Refs.\ \cite{Gaspard90,Delande93}: $P(c)\sim c^{-4}$.
In the same way for GSE, using $R_2(\epsilon)\sim\epsilon^4$, one obtains
the divergence of the moments starting from the {\em fifth} one, in agreement
with
$P(c)\sim c^{-6}$ \cite{Gaspard90,Delande93}. This is the reason, why the
result for the GOE curvature distribution, Eq.\ (\ref{goec}), coincides exactly
with that of Ref.\ \cite{Delande93}.

However, when $\alpha$ plays the role of a
magnetic flux, it does change the symmetry class of the Hamiltonian.
Then the energies $\epsilon_k(\alpha)$ are eigenvalues of the sum
$H_S^{(N)}+i\alpha H_A^{(N)}$, and thus no longer statistically
independent of the matrix elements, $<\alpha,k|H_A^{(N)}|n,\alpha>$.
We are
going to show  that in this case, for $\alpha\neq 0$, all the moments
converge. This in turn means that, contrary to the case of Refs.\
\cite{Gaspard90,Delande93}, the curvature DF has an exponential tail.
For  small $\alpha$ and small $|\epsilon_1-\epsilon_2|$ one can show that
$$
\left|\epsilon_1(\alpha)-\epsilon_2(\alpha)\right| \approx
\left((\epsilon_1-\epsilon_2)^2+\alpha^2 X^2 \right)^{1/2},
$$
where the quantity X depends only on the matrix elements (and not on the
energies),
hence it is statistically independent of the zero--flux energies,
$\epsilon_k$.
Repeating again the steps, leading to Eq.\ (\ref{lam}) one obtains
\begin{equation}
\langle c^m \rangle\sim \int f(X)dX \int_{0}^{\infty}
\frac{R_2(\epsilon)}{(\epsilon^2+\alpha^2 X^2)^{m/2}} d\epsilon \sim \left\{
\begin{array}{ll}
\mbox{const}; \hskip .5cm   & m=1, \\
\ln\alpha^{-1};            & m=2, \\
\alpha^{2-m};              & m\geq 3,
\end{array} \right.
                                                        \label{lam1}
\end{equation}
where $f(X)$ denotes the (unspecified) distribution of matrix elements. In
Eq.\ (\ref{lam1}) we used again $R_2(\epsilon)\sim\epsilon$.  Thus we
have shown that the small flux
behavior of the moments of the curvature, initially derived for
a $2\times 2$ model, Eq.\ (\ref{new}), is valid  for large $N$ as well.

\appendix{List of distributions for pure ensembles}
\label{s6}

{\em GOE}; $\alpha=0$

Spacing distribution:
\begin{equation}
P_0(\epsilon)=\frac{\epsilon}{4 v^2}
\exp\left(-\frac{\epsilon^2}{8v^2}\right),
                                                          \label{goe1}
\end{equation}
\begin{equation}
\langle\epsilon^m\rangle_0=v^m 2^{3m/2}\Gamma\left(\frac{m+2}{2}\right).
                                                          \label{goe2}
\end{equation}

Current distribution
\begin{equation}
P_0(i)=\frac{1}{v}2\delta\left(\frac{i}{v}\right),
                                                          \label{goe3}
\end{equation}
\begin{equation}
\langle i^m\rangle_0=0.
                                                          \label{goe4}
\end{equation}

Curvature distribution
\begin{equation}
P_0(c)=\frac{3}{8v}\sqrt{\frac{2v}{c}} \exp\left(\frac{1}{4}
\left(\frac{c}{2v}\right)^2\right) D_{-5/2}\left(\frac{c}{2v}\right),
                                                          \label{goe5}
\end{equation}
\begin{equation}
\langle c\rangle_0=v\sqrt{\frac{\pi}{2}}; \hskip 1 cm
\langle c^m\rangle_0=\infty; \hskip 1cm m\geq 2.
                                                          \label{goe6}
\end{equation}


{\em GUE}; $\alpha=1$

Spacing distribution:
\begin{equation}
P_1(\epsilon)=\frac{\epsilon^2}{4 \sqrt{2\pi}v^3}
\exp\left(-\frac{\epsilon^2}{8v^2}\right),
                                                          \label{gue1}
\end{equation}
\begin{equation}
\langle\epsilon^m\rangle_1=v^m \frac{2^{3m/2+1}}{\sqrt{\pi}}
\Gamma\left(\frac{m+3}{2}\right).
                                                          \label{gue2}
\end{equation}

Current distribution
\begin{equation}
P_1(i)=\frac{1}{v}\frac{1}{\sqrt{2\pi}}\sqrt{\frac{\sqrt{2}v}{i}}\,
\Gamma\left(\frac{5}{4},\left(\frac{i}{\sqrt{2}v}\right)^2\right),
                                                          \label{gue3}
\end{equation}
\begin{equation}
\langle i^m\rangle_1=v^m \frac{2^{m/2}}{\sqrt{\pi}}
\Gamma\left(\frac{m+3}{2}\right)\frac{1}{m+1/2}.
                                                          \label{gue4}
\end{equation}

Curvature distribution
\begin{equation}
P_1(c)\approx \frac{1}{v}\left\{ \begin{array}{ll}
{\displaystyle \frac{\Gamma(1/4)}{2^{9/4}\sqrt{\pi}}\, \sqrt{\frac{v}{c}};}
             & c\ll v, \\
{\displaystyle 4\frac{c}{v}\exp\left(-8\left(\frac{c}{v}\right)^2\right);}
\hskip .5 cm & v\ll c,
\end{array} \right.
                                                         \label{gue5}
\end{equation}
\begin{equation}
\langle c^m\rangle_1=v^m\frac{2^{m/2}}{\sqrt{\pi}}
\Gamma\left(\frac{m+3}{2}\right)
\sum_{p=0}^m\frac{(-1)^p\left(\stackrel{\textstyle m}{\textstyle p}\right)}
{m+1/2+p}.
                                                          \label{gue6}
\end{equation}

\end{document}